\documentstyle[12pt]{article}
\newcommand{\lmd}{\lambda}

\newcommand{\up}{\uparrow}
\newcommand{\down}{\downarrow}
\newcommand{\da}{\dagger}
\makeatletter
\renewcommand\thesection{\@Roman\c@section}
\renewcommand\thesubsection{\thesection.\@arabic\c@subsection}
\makeatother

\newcommand{\sect}[1]{\setcounter{equation}{0}\section{#1}}

\begin{document}
\begin{center}
{\Large \bf Supersymmetric $t$-$J$ Gaudin Models and KZ Equations}
\vskip.2in
{\large  Mark D.Gould $^{a}$, Yao-Zhong Zhang $^{a}$ and Shao-You Zhao
$^{b}$}
\vskip.2in
{\em $^a$ Center of Mathematical Physics, Department of mathematics, the
university of Queensland,
Brisbane 4072, Australia

 $^b$ P.O. Box 2324, Department of Physics, Nanjing University,
         Nanjing 210093, The People's Republic of China
}
\end{center}
\begin{abstract}
Supersymmetric $t$-$J$ Gaudin models with both periodic and open boundary
conditions are constructed and diagonalized by means of the algebraic 
Bethe ansatz method. Off-shell Bethe ansatz equations of the Gaudin
systems are derived, and used to construct and solve the KZ equations
associated with $sl(2|1)^{(1)}$ superalgebra.
\end{abstract}

\sect{introduction}

In the study of one dimensional long-range interacting systems,
Gaudin type models \cite{Gau76} ocupied an important place, due to
their role in establishing the integrability of the Seiberg-Witten
theory \cite{Sei94,Bra99} and diagonalizing the BCS hamiltonian of
ultrasmall metallic grains \cite{Ami01,Zhou02,Del01}. They also served 
as a testing ground
for ideas such as the functional Bethe ansatz and general procedure
of separation of variables \cite{Skl87,Skl96,Skl99}.

The $t$-$J$ model was proposed in an attempt to understand high-$T_c$
superconductivity \cite{And90}. It is a correlated electron system with
nearest-neighbor hopping ($t$) and anti-ferromagnetic exchange ($J$) of
electrons. Using the nested algebraic Bethe ansatz method, Essler and 
Korepin obtained the eigenvalues of the periodic system \cite{Ess92}. Soon
after the open boundary case was studied in \cite{Fan99,Gon94}. 

In \cite{Cao01}, the periodic $t$-$J$ Gaudin model was investigated and
its eigenvalues were obtained. In this paper, we study both the periodic
and open boundary $t$-$J$ Gaudin models by a method different from that
used in \cite{Cao01}.

The Knizhnik-Zamolodchikov (KZ) equations were first proposed as a set of
differential equations satisfied by correlation functions of
the Wess-Zumino-Witten models \cite{KZ}. The connection 
between Gaudin type magnets and the KZ equations has been studied by
many authors \cite{Babu94,Hik94,Has94,Fei94,Hik95}. We are interested in
the super KZ equations associated with $sl(2|1)^{(1)}$ superalgebra.
We will construct and solve these KZ equations with the help of the
$t$-$J$ Gaudin models.

The outline of this paper is as follows: In the first part of this paper 
(section 2-4), we study the periodic $t$-$J$ Gaudin model and its
corresponding KZ equation.  The $t$-$J$ model hamiltonian
is constructed in section 2, and
diagonalized in section 3, by using the algebraic Bethe ansatz method.
We also derive its off-shelled Bethe ansatz equations, and use them, in
section 4, to construct solution to the corresponding KZ equation.  
The second part of this paper is devoted to dealing with the open boundary
$t$-$J$ Gaudin model. In sections 5 and 6, we construct and diagonalize
the open boundary $t$-$J$ Gaudin model. In section 7, we obtain solution
to the KZ equation associated with the boundary Gaudin model.

\sect{Periodic $t$-$J$ Gaudin model}

\subsection{$t$-$J$ model with periodic boundary condition}

The supersymmetric $t$-$J$ model is described by the R-matrix 
arising from the 3-dimensional
representation of $sl(2|1)^{(1)}$. In the Fermionic, 
Fermionic and Bosonic (FFB) grading of the representation space, 
the R-matrix is given by \cite{Perk81}
\begin{equation}
R(\lmd)=\left( \begin{array}{ccccccccc}
a(\lmd)&0&0& 0&0&0& 0&0&0\\
0&b(\lmd)&0& -c_-(\lmd)&0&0& 0&0&0\\
0&0&b(\lmd)& 0&0&0& c_-(\lmd)&0&0\\
0&-c_+(\lmd)&0& b(\lmd)&0&0& 0&0&0\\
0&0&0& 0&a(\lmd)&0&  0&0&0\\
0&0&0& 0&0&b(\lmd)&  0&c_-(\lmd)&0\\
0&0&c_+(\lmd)& 0&0&0& b(\lmd)&0&0\\
0&0&0& 0&0&c_+(\lmd)& 0&b(\lmd)&0\\
0&0&0& 0&0&0& 0&0&w(\lmd)\end{array}\right),\label{R}
\end{equation}
where $\eta$ is a crossing parameter and
\begin{eqnarray}
&& a(\lmd)=1,\quad b(\lmd)=\frac{\sinh(\lmd)}{\sinh(\lmd-\eta)}, \quad
w(\lmd)=\frac{\sinh(\lmd+\eta)}{\sinh(\lmd-\eta)},\nonumber\\
&& c_+(\lmd) = \frac{e^{\lmd}\sinh(\eta)}{\sinh(\lmd-\eta)},\quad 
c_-(\lmd) =\frac{e^{-\lmd}\sinh(\eta)}{\sinh(\lmd-\eta)},
\end{eqnarray}
This $R$-matrix satisfies the graded Yang-Baxter equation (YBE)
\begin{equation}
R(\lambda -\mu )_{a_1a_2}^{b_1b_2}
R(\lambda )_{b_1a_3}^{c_1b_3}
R(\mu )_{b_2b_3}^{c_2c_3}
(-)^{(\epsilon _{b_1}+\epsilon _{c_1})\epsilon _{b_2}}
=
R(\mu )_{a_2a_3}^{b_2b_3}R(\lambda )_{a_1b_3}^{b_1c_3}
R(\lambda -\mu )_{b_1b_2}^{c_1c_2}(-)^{(\epsilon _{a_1}
+\epsilon _{b_1})\epsilon _{b_2}},
\end{equation}
where $\epsilon_a$ is the Grassman parity: $\epsilon_a=0$ for 
bosons and $\epsilon_a=1$ for fermions.
The R-matrix satisfies the unitarity and cross-unitarity relations,
\begin{eqnarray}
&&R_{12}(\lmd)R_{21}(-\lmd)=\rho(\lmd)\cdot id,\quad 
\rho(\lmd)=-\sinh(\lmd+\eta)\sinh(\lmd-\eta),\label{uni}\nonumber\\
&&R^{st_1}_{12}(\lmd-\eta)M_1R^{str_1}_{21}M^{-1}_1
        =\tilde\rho(\lmd)\cdot id,\quad
\tilde\rho(\lmd)=\sinh(\lmd)\sinh(\lmd-\eta),\label{cross-uni}
\end{eqnarray}
where $M$ is a diagonal matrix diag$(e^{2\eta},1,1)$ and  $st$ is the 
super-transposition defined by 
\begin{equation}
(A^{st})_{ij}=A_{ji}(-1)^{(\epsilon_i+1)\epsilon_j}.
\end{equation}

Consider the L-operator 
\begin{eqnarray}
L_{aq}(\lambda )\equiv R_{aq}(\lambda ),
\end{eqnarray}
where $a$ represents the auxiliary space and $q$
represents the quantum space. The L-operator also obeys
the (graded) YBE
\begin{eqnarray}
R_{12}(\lambda -\mu )L_1(\lambda )L_2(\mu )
=L_2(\mu )L_1(\lambda )R_{12}(\lambda -\mu ). 
\end{eqnarray}
The tensor product is graded, namely, 
\begin{eqnarray}
(F\otimes G)_{ac}^{bd}=F_a^bG_c^d(-)^{(\epsilon _a+\epsilon _b)\epsilon
_c}.
\end{eqnarray}
The row-to-row monodromy matrix
$T_N(\lambda )$ is defined as the product of $N$ operators,
\begin{eqnarray}
T_a(\lambda )=L_{a1}(\lmd-z_1)L_{a2}(\lmd-z_2)\cdots L_{aN}(\lambda-z_N
),\label{T-p}
\end{eqnarray}
In matrix form,
\begin{eqnarray}
\{ [T(\lambda )]^{ab}\}^{ \alpha _1\cdots \alpha _N}_{
\beta _1\cdots \beta _N}
&=&L_1(\lambda-z_1)_{a\alpha _1}^{c_1\beta _1}
L_{2}(\lambda-z_2)_{c_1\alpha _{2}}^{c_{2}\beta _{2}}
\cdots L_N(\lambda-z_N)_{c_{N-1}\alpha _N}^{b\beta _N}\nonumber\\
& &(-1)^{\sum _{j=1}^{N-1}(\epsilon _{\alpha _j}+\epsilon _{\beta _j})
\sum _{i=j+1}^{N}\epsilon _{\alpha _i}}\nonumber\\
\end{eqnarray}  
By repeatedly using the YBE, one can easily check that
the monodromy matrix satisfies 
\begin{eqnarray}
R(\lambda -\mu )T_1(\lambda )T_2(\mu )
=T_2(\mu )T_1(\lambda )R(\lambda -\mu ). \label{YBR}
\end{eqnarray}

The transfer matrix $t(\lambda )$ is defined as the supertrace of the
monodromy matrix over the auxiliary space:
\begin{eqnarray}  
t(\lambda )=strT(\lambda )
=\sum (-1)^{\epsilon _a}T(\lambda )_{aa}.\label{transfer-p}
\end{eqnarray} 
Using the YBE, one can show that the transfer matrix $t(\lmd)$ constitutes
a one-parameter commuting family, i.e.
\begin{equation}
[t(\lmd),t(\mu)]=0.
\end{equation}
Therefore, the $t$-$J$ model is integrable.

\subsection{Supersymmetric $t$-$J$ Gaudin model}

Supersymmetric $t$-$J$ Gaudin model can be obtained by taking the 
quasi-classical limit $\eta\rightarrow 0$ of the transfer matrix at 
the point $\lmd=z_j$
\cite{Hik92}. So we expand the R-matrix, L-operator and 
transfer matrix around the point $\eta=0$ to get
\begin{eqnarray}
&&R(\lambda)=1+\eta\hat R(\lmd)+{\cal O}(\eta^2),                                     \label{limit-R}\\
&&L(\lambda)=1+\eta{\cal L}(\lmd)+{\cal O}(\eta^2),
                    \label{limit-L} \\
&&t(z_j)=-1+\eta\hat t(z_j)+{\cal O}(\eta^2).
                     \label{limit-t}
\end{eqnarray}
Then the Hamiltonian of periodic $t$-$J$ Gaudin model can be obtained 
from the second term of (\ref{limit-t}).  

 From (\ref{limit-R}), we have
\begin{eqnarray}
\hat R_{ij}(\lmd)&=&2\coth(\lmd)e^i_{33}\stackrel{s}{\otimes}e^j_{33}
+\coth(\lmd)\left[e^i_{11}\stackrel{s}{\otimes}e^j_{22}
   +e^i_{11}\stackrel{s}{\otimes}e^j_{33}\right.\nonumber\\&&\mbox{}\left.
   +e^i_{22}\stackrel{s}{\otimes}e^j_{11}
   +e^i_{33}\stackrel{s}{\otimes}e^j_{11}
   +e^i_{33}\stackrel{s}{\otimes}e^j_{22}\right]\nonumber\\&&\mbox{}
+e^{-\lmd}\frac{1}{\sinh(\lmd)}
   \left[e^i_{13}\stackrel{s}{\otimes}e^j_{31}
        +e^i_{23}\stackrel{s}{\otimes}e^j_{32}
        -e^i_{12}\stackrel{s}{\otimes}e^j_{21}\right]\nonumber\\&&\mbox{}
+e^{\lmd}\frac{1}{\sinh(\lmd)}
   \left[e^i_{31}\stackrel{s}{\otimes}e^j_{13}
        +e^i_{32}\stackrel{s}{\otimes}e^j_{23}
        -e^i_{21}\stackrel{s}{\otimes}e^j_{12}\right],
\end{eqnarray}
where $e^k_{ij}$ is a matrix acting on the $k$-th space with
elements $(e_{ij})_{\alpha\beta}=\delta_{i\alpha}\delta_{j\beta}$.
Denote by $S,S^{\dagger}, S^z, Q_{\pm 1},
Q_{\pm 1}^{\dagger}$ and $T^z$
the generators of  $sl(2|1)$, which satisfy, among others,
\begin{equation}
[S^\dagger, S]=S^z,~~~~\{Q^\dagger_{1},Q_{1}\}=T^z,~~~~\{Q^\dagger_{-1},
   Q_{-1}\}=S^z+T^z.
\end{equation}
In the fundamental representation, they
take the form,
\begin{eqnarray}
&&S^z=\left(\begin{array}{ccc}
1&0&0\\ 0&-1&0\\ 0&0&0 \end{array}\right),\quad
  T^z=\left(\begin{array}{ccc}
0&0&0\\ 0&1&0\\ 0&0&1 \end{array}\right), \nonumber\\
&& S=e_{21},\quad S^{\dagger}=e_{12},\quad
 Q_{1}=e_{32},\nonumber\\
&&Q_{1}^{\dagger}=e_{23},\quad Q_{-1}=e_{31},\quad
Q_{-1}^{\dagger}=e_{13}.
\end{eqnarray}
Thus, 
\begin{eqnarray}
\hat R_{ij}(\lmd)&=&2\coth(\lmd)\left[(1-(S^z_i)^2)(1-(S^z_j)^2)\right]
\nonumber\\&&\mbox{}
+\coth(\lmd)\left[(1-T_i^z)(2-T_j^z-S^z_j-(S^z_j)^2)
    +(2-T_i^z-S^z_i-(S^z_i)^2)(1-T_j^z)\right.\nonumber\\ &&\mbox{}\left.
    +(1-T_i^z-S^z_i)(1-(S^z_j)^2)
    +(1-(S^z_i)^2)(1-T_j^z-S^z_j)\right]
\nonumber\\ &&\mbox{}
+\frac{e^{-\lmd}}{\sinh(\lmd)}\left[
    -S^\da_iS_j+\sum_{\sigma=\up,\down}Q^\da_{i,\sigma}Q_{i,\sigma}\right]
%\nonumber\\ &&\mbox{}
+\frac{e^{\lmd}}{\sinh(\lmd)}\left[
-S_iS^\da_j+\sum_{\sigma=\up,\down}Q_{i,\sigma}Q^\da_{i,\sigma}\right]
\nonumber\\
\end{eqnarray}
Using the standard fermionic representation \cite{Ess92}   
\begin{eqnarray}
&&S_j=c^{\dagger}_{j,1}c_{j,-1},\quad
S_j^{\dagger}=c^{\dagger}_{j,-1}c_{j,1},
\quad S^z_j=n_{j,-1}-n_{j,1},\nonumber\\
&&T^z_j=1-n_{j,-1},\quad
Q^{\dagger}_{j,\sigma}=c^{\dagger}_{j,\sigma}(1-n_{j,-\sigma}),\quad
 Q_{j,\sigma}=c_{j,\sigma}(1-n_{j,-\sigma}),\quad
\end{eqnarray}
where $n_{j,\pm}=c^\dagger_{j,\pm}c_{j,\pm}$ and $n_{j,+}+n_{j,-}=n_j$,
we have
\begin{eqnarray}
\hat R_{ij}(\lmd)&=&2\coth(\lmd)\left[(1-(n_{i,\down}-n_{i,\up})^2)
                           (1-(n_{j,\down}-n_{j,\up})^2)\right]
\nonumber\\&&\mbox{}
+\coth(\lmd)\left[n_{i,\down}
      (1+n_{j,\up}-(n_{j,\down}-n_{j,\up})^2)
      +(1+n_{i,\up}-(n_{i,\down}-n_{i,\up})^2)n_{j,\down}
\right.\nonumber\\&&\mbox{}\left.
      +n_{i,\up}(1-(n_{i,\down}-n_{i,\up})^2)
      +(1-(n_{i,\down}-n_{i,\up})^2)n_{j,\up}\right]\nonumber\\&&\mbox{}
+\frac{e^{-\lmd}}{\sinh(\lmd)}\left[
   -c^\da_{i,\down}c_{i,_\up}c^\da_{j,\up}c_{j,\down}
   +\sum_{\sigma=\up,\down}
    c^\da_{i,\sigma}(1-n_{i,-\sigma})
    c_{j,\sigma}(1-n_{j,-\sigma})\right]\nonumber\\&&\mbox{}
+\frac{e^{\lmd}}{\sinh(\lmd)}\left[
   -c^\da_{i,\up}c_{i,_\down}c^\da_{j,\down}c_{j,\up}
   +\sum_{\sigma=\up,\down}
    c_{i,\sigma}(1-n_{i,-\sigma})
    c^\da_{j,\sigma}(1-n_{j,-\sigma})\right].
\end{eqnarray}

Similarly, from the eq.(\ref{limit-L}), we obtain the elements of the
Lax matrix,
\begin{eqnarray}
&&{\cal L}^j_{11}=\coth(\lmd)-\coth(\lmd)n_{j,-1},\quad
{\cal L}^j_{22}=\coth(\lmd)-\coth(\lmd)n_{j,1},\nonumber\\
&&{\cal L}^j_{33}=\coth(\lmd)
   +\coth(\lmd)\left[1-(n_{j,-1}-n_{j,1})^2\right], \nonumber\\
&&{\cal L}^j_{12}=-{e^{-\lmd}\over\sinh(\lmd)}c^+_{j,1}c_{j,-1}\quad
{\cal L}^j_{21}=-{e^{\lmd}\over\sinh(\lmd)}c^+_{j,-1}c_{j,1},
           \nonumber\\
&&{\cal L}^j_{13}={e^{-\lmd}\over\sinh(\lmd)}c_{j,-1}(1-n_{j,1})\quad
{\cal L}^j_{31}={e^{\lmd}\over\sinh(\lmd)}c^+_{j,-1}(1-n_{j,1}),
           \nonumber\\
&&{\cal L}^j_{23}={e^{-\lmd}\over\sinh(\lmd)}c_{j,1}(1-n_{j,-1})\quad
{\cal L}^j_{32}={e^{\lmd}\over\sinh(\lmd)}c^+_{j,-1}(1-n_{j,1}),
\end{eqnarray}

Finally, the Hamiltonian of the $t$-$J$ Gaudin model can be written as
\begin{eqnarray}
H_j&=&\frac{dt(\lmd=z_j)}{d\eta}|_{\eta=0}.\nonumber\\
&=&\sum_{k=1\ne j}^N\frac{1}{\sinh(z_j-z_k)}
   \left\{ \cosh(z_j-z_k)[-n_{j,-1}n_{k,-1}-n_{j,1}n_{k,1}+(1-n_j)(1-n_k)]
  \right.\nonumber\\
&&\quad\quad\mbox{}+\ e^{-(z_j-z_k)}\left[\sum_{\sigma=\pm1}
c^{\dagger}_{j,\sigma}(1-n_{j,-\sigma})c_{k,\sigma}(1-n_{k,-\sigma})
-S^{\dagger}_{j}S_{k}\right]\nonumber\\
&&\quad\quad\mbox{}\left.+\ e^{z_j-z_k}\left[\sum_{\sigma=\pm1}
c_{j,\sigma}(1-n_{j,-\sigma})c^{\dagger}_{k,\sigma}(1-n_{k,-\sigma})
-S_{j}S^{\dagger}_{k}\right]\right\}.\label{ham-p}
\end{eqnarray}

\sect{Bethe ansatz for the $t$-$J$ Gaudin model}

Hamiltonian (\ref{ham-p}) can be diagonalized by using the algebraic
Bethe ansatz method.  To simplify calculation, we use the 
gauge transformation 
\begin{equation}
R(\lmd)\rightarrow {\rm diag}
  \left(e^{\lmd/2},e^{\lmd/2},e^{-\lmd/2}\right)\stackrel{s}{\otimes}1
\cdot R(\lmd)\cdot
 {\rm diag} \left(e^{-\lmd/2},e^{-\lmd/2},e^{\lmd/2}\right)
\stackrel{s}{\otimes}1
\end{equation}
to get
\begin{equation}
R(\lmd)=\left( \begin{array}{ccccccccc}
a(\lmd)&0&0& 0&0&0& 0&0&0\\
0&b(\lmd)&0& -c_-(\lmd)&0&0& 0&0&0\\
0&0&b(\lmd)& 0&0&0& d(\lmd)&0&0\\ 
0&-c_+(\lmd)&0& b(\lmd)&0&0& 0&0&0\\
0&0&0& 0&a(\lmd)&0&  0&0&0\\
0&0&0& 0&0&b(\lmd)&  0&d(\lmd)&0\\
0&0&d(\lmd)& 0&0&0& b(\lmd)&0&0\\
0&0&0& 0&0&d(\lmd)& 0&b(\lmd)&0\\
0&0&0& 0&0&0& 0&0&w(\lmd)\end{array}\right),\label{mR}
\end{equation}
where $d(\lmd)=\sinh(\eta)/\sinh(\lmd-\eta)$.
The corresponding $L$-matrix can be written as
\begin{eqnarray} &&
L_n(\lambda )\nonumber\\
&=&\left(
\begin{array}{ccc}
b(\lambda )-(b(\lambda )-a(\lambda ))e^n_{11}&
-c_-(\lambda )e^n_{21} &d(\lambda )e^n_{31}\\ 
-c_+(\lambda )e^n_{12} &b(\lambda )-(b(\lambda )-a(\lambda ))e^n_{22}&
d(\lambda )e^n_{32}\\
d(\lambda )e^n_{13} &d(\lambda )e^n_{23} &
b(\lambda )-(b(\lambda )-w(\lambda ))e^n_{33}
\end{array}\right).\nonumber\\
\end{eqnarray}

We expand the row-to-row monodromy matrix ({\ref{T-p}) around $\eta=0$:
\begin{eqnarray}
T(\lambda )&=&\left( \begin{array}{ccc}
A_{11}(\lambda )&A_{12}(\lambda ) &B_1(\lambda )\\
A_{21}(\lambda )&A_{22}(\lambda )&B_2(\lambda )\\
C_1(\lambda )&C_2(\lambda )&D(\lambda )\end{array}
\right)\nonumber\\[3mm]
&=&1+\eta\hat T(\lmd)+{\cal O}(\eta^2)\nonumber\\[3mm]
&=&1+\eta
\left( \begin{array}{ccc}
\hat A_{11}(\lambda )&\hat A_{12}(\lambda ) &\hat B_1(\lambda )\\
\hat A_{21}(\lambda )&\hat A_{22}(\lambda )&\hat B_2(\lambda )\\
\hat C_1(\lambda )&\hat C_2(\lambda )&\hat D(\lambda )\end{array}
\right)+{\cal O}(\eta^2).
\end{eqnarray}

By the graded YBE (\ref{YBR}), one finds
\begin{eqnarray}
[\hat T_1(\lmd),\hat T_2(\mu)]
=[\hat T_1(\lmd)+\hat T_2(\mu),\hat R_{12}(\lmd-\mu)], \label{YBR-c}
\end{eqnarray}
and the commutation relations, 
\begin{eqnarray}
\hat C_{s_1}(\mu_1)\hat C_{s_2}(\mu_2)
      &=&-\hat C_{s_2}(\mu_2)\hat C_{s_1}(\mu_1),\label{commu-C2}\\[3mm]
\hat D(z_j)\hat C_s(\mu)
   &=&-\coth(z_j-\mu)\hat C_s(\mu)D(z_j)|_{\eta=0} 
     +\hat C_s(\mu)\hat D(z_j)\nonumber\\
&&\mbox{}     
     +\frac{1}{\sinh(\mu-z_j)}E^-_j(s)\hat D(\mu)\nonumber\\
&&\mbox{}    
      -\frac{1}{\sinh(\mu-z_j)}\hat C_{s}(z_j) D(\mu)_{\eta=0},
                 \label{commu-D} \\[3mm]
\hat A_{s_1s_2}(z_j)\hat C_{p_1}(\mu)
   &=&\left(r^{s_2p_1}_{MN}(z_j-\mu)\right)'_{\eta=0}
           \hat C_s(\mu)A_{s_1m}(z_j)_{\eta=0}
     +\hat C_{p_1}(\mu)\hat A_{s_1s_2}(z_j)\label{commu-A}\nonumber\\
&&\mbox{}    
     +\frac{1}{\sinh(\mu-z_j)}E^-_j(s_2)\hat A_{s_1p_1}(\mu)\nonumber\\
&&\mbox{}    
     -\frac{1}{\sinh(\mu-z_j)}
        \hat C_{s_2}(z_j)A_{s_1p_1}(\mu)_{\eta=0},\nonumber\\ 
\end{eqnarray}
where $j$ indicates the lattice position and $E^-_j(s)$ acts on the 
quantum space with $E^-_j(s)=e_{13}$ for $s=1$ and
$E^-_j(s)=e_{23}$ for $s=2$; the r-matrix $r(\lambda)$ is defined by
\begin{equation}
r(\lmd)={1\over\sinh(\lmd)}\left(\begin{array}{cccc}
\sinh(\lmd-\eta)&0&0&0\\
0&\sinh(\lmd)&-e^{-\lmd}\sinh(\eta)&0\\
0&-e^{\lmd}\sinh(\eta)&\sinh(\lmd)&0\\
0&0&0&\sinh(\lmd-\eta)\end{array}\right)
\end{equation}
Define the vacuum state:
\begin{eqnarray}
|0>_n=\left(\begin{array}{c}
0\\0\\1\end{array}\right),~~~~
|0>=\otimes _{k=1}^N|0>_k.\label{vacuum}
\end{eqnarray}
Then, 
\begin{eqnarray}
&&\hat B_a(\lambda )|0>=0,
~~~\hat C_a(\lambda )|0>\not= 0,
\nonumber \\
&&\hat D(\lambda)|0>=\sum_{i=1}^N2\coth(\lambda-z_i)|0>\nonumber \\
&&\hat A_{ab}(\lambda )|0>=\left\{\begin{array}{l}
0\quad \mbox{for } \lmd=z_j,\mbox{ and }a\ne b \\ 
\sum_{i=1}^N \coth(\lmd-z_i) \quad
  \mbox{for }\lmd\ne z_j, \mbox{ and }a=b 
\end{array}\right.. 
\label{eigenv-p}
\end{eqnarray}
where $j=1,2,\cdots,N.$

Define the Bethe state 
\begin{eqnarray}
\phi=\hat C_{d_1}(\mu _1)
\hat C_{d_2}(\mu _2)\cdots \hat C_{d_n}(\mu _n)|0>F^{d_1\cdots
d_n},
\end{eqnarray}  
where $F^{d_1\cdots d_n}$ is a function of the spectral parameters $\mu
_j$. We moreover define state
\begin{equation}
\phi^{(1)}=\hat C^{(1)}(\mu^{(1)}_1)\hat C^{(1)}(\mu^{(1)}_2)\cdots
           \hat C^{(1)}(\mu^{(1)}_m).
\end{equation}
Then $\phi^{(1)}$ spans a subspace of the space spanned by
$\phi$. Applying the transfer matrix to these states and keeping in
mind of the commutation relation (\ref{commu-D}), we can find the
eigenvalues $E_j$ of $t(z_j)$ and the
Bethe ansatz equations. This is done as follows. Firstly,
from (\ref{eigenv-p}) we have  
\begin{eqnarray}
\hat D(z_j)\phi&=&\left[\sum_{\alpha=1}^n\coth(z_j-\mu_\alpha)
      -2\sum_{k=1,\ne j}^N \coth(z_j-z_k)\right]\phi
\nonumber\\&& \mbox{}
  +\sum_{\alpha=1}^n\frac{(-1)^{\alpha-1}}{\sinh(\mu_\alpha-z_j)}
   \left[2\sum_{k=1}^N\coth(\mu_\alpha-z_k)
\right.\nonumber\\ &&\mbox{}\left.
  -\sum_{\beta=1,\ne \alpha}^n
    \coth(\mu_\alpha-\mu_\beta)\right]E^-_j(d_\alpha)\phi_\alpha 
           \nonumber\\&&\mbox{}
  -\sum_{\alpha=1}^n\frac{(-1)^{\alpha-1}}
    {\sinh(\mu_\alpha-z_j)}\hat C_{d_\alpha}(z_j)\phi_\alpha,
                \label{commu-DB}\nonumber\\
\end{eqnarray}
where $\phi_\alpha=\prod_{\beta=1,\ne\alpha}^n \hat
C_{d_\beta}(\mu_\beta)$. Secondly, the action of $\hat A(z_j)$ 
on $\phi$ is given by
\begin{eqnarray}
-\hat A_{aa}(z_j)\phi&=&
\sum_{\alpha=1}^n
\frac{(-1)^{\alpha-1}}{\sinh(\mu_\alpha-z_j)}
\left[-\sum_{k=1}^N\coth(\mu_\alpha-z_k)
\right.%\nonumber\\ &&\mbox{} 
+\left.       \Lambda^{(1)}(\mu_\alpha)\right]
       E^-_j(d_\alpha)\phi_\alpha  \nonumber\\&&\mbox{}
   +\sum_{\alpha=1}^n\frac{(-1)^{\alpha-1}}
    {\sinh(\mu_\alpha-z_j)}\hat C_{d_\alpha}(z_j)\phi_\alpha,
         \label{commu-AB} \nonumber\\
\end{eqnarray}
where $\Lambda^{(1)}(\mu_\alpha)$ is the operator
\begin{eqnarray}
t^{(1)}(\lmd)&=&{d\over d\eta}
str\left[ r^{a~d_1}_{m_1n_1}(\lmd-\mu_1)
  r^{m_1d_2}_{m_2n_2}(\lmd-\mu_2)
  \cdots\right.	% \nonumber\\&&\mbox{}
\left.
  r^{m_{n-1}d_M}_{m_n~~n_n}(\lmd-\mu_n)
      \right]_{\eta=0}.\nonumber\\
\end{eqnarray}
Using results in the appendix, we obtain
\begin{eqnarray}
\Lambda^{(1)}(\mu_\alpha)
 =\sum_{\beta=1,\ne\alpha}^n\coth(\mu_\alpha-\mu_\beta)
  -\sum_{\gamma=1}^m\coth(\mu_\alpha-\mu_\gamma^{(1)}),\label{eigenv-n} 
\end{eqnarray}
where $\mu^{(1)}_1,\cdots, \mu^{(1)}_m$ satisfy the constraints
\begin{eqnarray}
f^{(1)}_\gamma\equiv 
   \sum_{\beta=1,\ne\alpha}^n\coth(\mu_\beta-\mu^{(1)}_\gamma)
   +2\sum_{\delta=1,\ne\gamma}^m
      \coth(\mu^{(1)}_\gamma-\mu^{(1)}_\delta)=0.\label{ba-n}
                      \nonumber\\
\end{eqnarray}

Thus, from (\ref{ham-p}),({\ref{commu-DB})-(\ref{commu-AB})
and ({\ref{eigenv-n})-(\ref{ba-n}),  we obtain the off-shell Bethe
ansatz equations
\begin{eqnarray}
H_j\phi&=&E_j\phi+\sum_{\alpha=1}^M 
  \frac{(-1)^{\alpha-1}}{\sinh(\mu_\alpha-z_j)}
       f_\alpha E^-_j(d_\alpha)\phi_\alpha,
\end{eqnarray}
where $\mu_\alpha$ satisfy the condition $f^{(1)}\gamma = 0$ and
\begin{eqnarray}
E_j&=&\sum_{\alpha=1}^n\coth(z_j-\mu_\alpha)-2\sum_{k=1,\ne j}^N
 \coth(z_j-z_k)\\
f_\alpha&=&-\sum_{\gamma=1}^m\coth(\mu_\alpha-\mu^{(1)}_\gamma)
+\sum_{k=1}^N\coth(\mu_\alpha-z_k),\\
\phi&=&\prod_{\alpha=1}^n\left(\sum_{k=1}^N{1\over\sinh(\mu_\alpha-z_k)}
           E^-_k(d_\alpha)\right)|0>F^{d_1\cdots d_n}\\
\phi_\alpha&=&
\prod_{\beta=1,\ne\alpha}^n\left(\sum_{k=1}^N{1\over\sinh(\mu_\beta-z_k)}  
           E^-_k(d_\beta)\right)|0>F^{d_1\cdots d_n}.
\end{eqnarray}

\sect{Super KZ equation}

As a set of partial differential equations,  the KZ equations take the form
\begin{equation}
\nabla_j\Psi=0 \quad \quad \mbox{for}\ \  j=1,2,\cdots,N, \label{kz-elli}
\end{equation}
where the differential operator $\nabla_j$  is defined by the Gaudin
Hamiltonian $H_j$:
\begin{equation}
\nabla_j=\kappa\frac{\partial}{\partial z_j}-H_j \label{elli-nabla}
\end{equation}
with $\kappa$ being a parameter.  Substituting (\ref{ham-p})
into (\ref{elli-nabla}), we can check
\begin{equation}
[\nabla_j,\nabla_k]=0,
\end{equation}
which ensures the integrability of the KZ equation.

To simplify our calculation, we make the following transformation
\begin{eqnarray*}
&&H_j\rightarrow H_j+
2\sum_{k=1,\ne j}^N\coth(z_j-z_k),\\
&&E_j\rightarrow E_j+2\sum_{k=1,\ne
j}^N\coth(z_j-z_k)=\sum_{\alpha=1}^n\coth(z_j-\mu_\alpha).
\end{eqnarray*}
Under this transformation, the form of the
off-shell Bethe ansatz equations is invariant. 

The function $\Psi(z)$ can be constructed by the hypergeometric function
$\chi(z,\mu)$ which oebys the equations
\begin{eqnarray}
\kappa\frac{\partial}{\partial z_j}\chi&=&E_j\chi, \nonumber\\
\kappa\frac{\partial}{\partial \mu_\alpha}\chi&=&f_\alpha\chi,
\end{eqnarray}
and the constraint $f^{(1)}_\gamma = 0$.
The solution to the above equations is given by
\begin{equation}
\chi(z,\mu)=\prod_{\beta<\alpha}
    [\sinh(\mu_\alpha-\mu^{(1)}_\gamma)]^{-1/\kappa}
            \prod_{\alpha=1}^n\prod_{j=1}^N
    [\sinh(z_j-\mu_\alpha)]^{1/\kappa}
\end{equation}
with $\mu_\alpha$ satisfying the condition $f_\gamma^{(1)}=0$.
With the help of $\chi(z,\mu)$, the function $\Psi(z)$ is given by
\begin{equation}
\Psi(z)=\oint_C\cdots\oint_C d\mu_1\cdots d\mu_n \chi(t,z)\phi(t,z),
\end{equation}  
where the integration path $C$ is a closed contour in the
Riemann surface such that the integrand resumes its initial value after
$t_\alpha$ has described it.
Substituting the expressions of $\nabla_j$ and $\Psi(z)$ into
({\ref{kz-elli}), we can show that the KZ equation is satisfied.
The proof is as follows
\begin{eqnarray}
\kappa\frac{\partial}{\partial z_j}\Psi(z)&=&
   \oint_C\cdots\oint_C d\mu_1\cdots d\mu_n \left(
       \kappa\frac{\partial \chi}{\partial z_j}\phi
       +\kappa\chi\frac{\partial \phi}{\partial z_j}\right)\nonumber\\ 
&=& \oint_C\cdots\oint_C dt_1\cdots d\mu_n \left(
       \chi E_j\phi
       +\kappa\chi\frac{\partial \phi}{\partial z_j}\right)\nonumber\\
&=&\oint_C\cdots\oint_C d\mu_1\cdots d\mu_n \left[
        \chi  H_j\phi-\chi\sum_{\alpha=1}^n
    W(\mu_\alpha,z_j)f_\alpha E^-_j(d_\alpha)\phi_\alpha
\right.\nonumber\\
&&        -\left.\kappa\chi\sum_\alpha
\frac{\partial }{\partial \mu_\alpha}\left(
W(\mu_\alpha,z_j)\phi_\alpha E_j^-({d_\alpha})\right)\right]
                                       \nonumber\\ 
&=&\oint_C\cdots\oint_C d\mu_1\cdots d\mu_n \left[
        \chi H_j\phi\right.\nonumber\\
&&        -\left.\kappa\sum_\alpha
         \frac{\partial }{\partial \mu_\alpha}\left(\chi
W(\mu_\alpha,z_j)
\phi_\alpha E_j^-({d_\alpha})\right)\right]
                                       \nonumber\\
&=&H_j\Psi,
\end{eqnarray}
where $W(\mu_\alpha,z)=(-1)^{\alpha-1}/\sinh(\mu_\alpha-z).$

\sect{Open boundary $t$-$J$ Gaudin model}

In this and the next section, we discuss the open boundary $t$-$J$ Gaudin 
system. We start with the graded reflection relation \cite{Fan99},
\begin{eqnarray}
&&R(\lmd-\mu)^{b_1b_2}_{a_1a_2}K(\lmd)^{c_1}_{b_1}
  R(\lmd+\mu)^{c_2d_1}_{b_2c_1}K(\mu)^{d_2}_{c_2}
  (-1)^{(\epsilon_{b_1}+\epsilon_{c_1})\epsilon_{b_2}}\nonumber\\
&&=K(\mu)^{b_2}_{a_2}R(\lmd+\mu)^{b_1c_2}_{a_1b_2}
     K(\lmd)^{c_1}_{b_1}R(\lmd-\mu)^{d_2d_1}_{c_2c_1}
      (-1)^{(\epsilon_{b_1}+\epsilon_{c_1})\epsilon_{c_2}}, \label{K}
\end{eqnarray}
where the $K(\lmd)$ is the reflection K-matrix. The diagonal solutions of
the reflection equation were found in \cite{Fan99}. In the present paper,
we only consider a special case in which $K(\lmd)=1$.

{}Following the standard procedure, we define the double-row monodromy 
matrix 
\begin{eqnarray}
{\cal T}(\lmd)=T(\lmd)K(\lmd)T^{-1}(-\lmd).\label{T-O}
\end{eqnarray}
Here $T(\lmd)$ is same as in the periodic case. 
One can check  that the following relation is satisfied:
\begin{eqnarray} 
&&R(\lmd-\mu)^{b_1b_2}_{a_1a_2}{\cal T}(\lmd)^{c_1}_{b_1}
  R(\lmd+\mu)^{c_2d_1}_{b_2c_1}{\cal T}(\mu)^{d_2}_{c_2}
  (-1)^{(\epsilon_{b_1}+\epsilon_{c_1})\epsilon_{b_2}}\nonumber\\
&&={\cal T}(\mu)^{b_2}_{a_2}R(\lmd+\mu)^{b_1c_2}_{a_1b_2}
     {\cal T}(\lmd)^{c_1}_{b_1}R(\lmd-\mu)^{d_2d_1}_{c_2c_1}
      (-1)^{(\epsilon_{b_1}+\epsilon_{c_1})\epsilon_{c_2}}. \label{Reflection}
\end{eqnarray}
                                                  
The dual reflection relation reads
\begin{eqnarray} 
&&R_{12}(\mu-\lmd)K_1^+(\lmd)M^{-1}_1
  R_{21}(\eta-\lmd-\mu)K_2^+(\mu)M^{-1}_2 \nonumber\\
&&=K_2^+(\mu)M_2^{-1}R_{12}(\eta-\lmd-\mu)
     K_1^+(\lmd)M^{-1}_1R_{21}(\mu-\lmd),
  \end{eqnarray}
where $M$ is a diagonal matrix $M={\rm diag}(e^{2\eta},1,1)$. Solution
$K^+$ associated with $K(\lambda)=1$ is given by
\begin{equation}
K^+(\lmd)\equiv MK(-\lmd+\eta/2)={\rm diag}(e^{2\eta},1,1).
\end{equation}

Define the boundary transfer matrix,
\begin{equation}
t^b(\lmd)=strK^+(\lmd){\cal T}(\lmd).\label{t-o}
\end{equation}
By (\ref{cross-uni}) and  (\ref{Reflection}), one can show the 
commutativity of the transfer matrix for different $\lambda$ values.

Similar to the periodic case, the boundary $t$-$J$ Gaudin system can  be 
obtained by expanding the boundary transfer matrix at the point $\lmd=z_j$
around $\eta=0$:
\begin{eqnarray}
t^b(\lmd=z_j)=1+\eta H^b_j+{\cal O}(\eta^2).
\end{eqnarray}
The second term on the right hand side gives the Hamiltonian of the 
open boundary $t$-$J$ Gaudin model. Explicitly, 
\begin{eqnarray}
H_j&=&{dt(z_j)\over d\eta}|_{\eta=0}\nonumber\\
&=&2(1+3\coth(2z_j))n_{j}
 \nonumber\\
&&\mbox{}
+\sum_{k=1\ne j}^N\frac{1}{\sinh(z_j-z_k)}
   \left\{ \cosh(z_j-z_k)[-n_{j,-1}n_{k,-1}-n_{j,1}n_{k,1}+(1-n_j)(1-n_k)]
  \right.\nonumber\\
&&\quad\quad\mbox{}+\ e^{-(z_j-z_k)}\left[\sum_{\sigma=\pm1}
c^{\dagger}_{j,\sigma}(1-n_{j,-\sigma})c_{k,\sigma}(1-n_{k,-\sigma})
-S^{\dagger}_{j}S_{k}\right]\nonumber\\
&&\quad\quad\mbox{}\left.+\ e^{z_j-z_k}\left[\sum_{\sigma=\pm1}
c_{j,\sigma}(1-n_{j,-\sigma})c^{\dagger}_{k,\sigma}(1-n_{k,-\sigma})
-S_{j}S^{\dagger}_{k}\right]\right\}\nonumber\\
&&+\sum_{k=1\ne j}^N \frac{1}{\sinh(z_j+z_k)}
   \left\{\cosh(z_j+z_k)[[-n_{j,-1}n_{k,-1}-n_{j,1}n_{k,1}+(1-n_j)(1-n_k)]
    \right.\nonumber\\
&&\quad\quad\mbox{}-\ e^{-(z_j+z_k)}\left[\sum_{\sigma=\pm1}
c_{j,\sigma}(1-n_{j,-\sigma})c^{\dagger}_{k,\sigma}(1-n_{k,-\sigma})
+S_{j}S^{\dagger}_{k}\right]\nonumber\\
&&\quad\quad\mbox{}\left.-\ e^{z_j+z_k}\left[\sum_{\sigma=\pm1}
c^{\dagger}_{j,\sigma}(1-n_{j,-\sigma})c_{k,\sigma}(1-n_{k,-\sigma})
+S^{\dagger}_{j}S_{k}\right]\right\}.
\end{eqnarray}

\sect{Bethe ansatz for boundary $t$-$J$ Gaudin model}

As in the periodic case,  we write the double-monodromy matrix as 
\begin{eqnarray}
{\cal T}(\lmd)=\left(\begin{array}{ccc}
{\cal A}_{11}(\lmd)& {\cal A}_{12}(\lmd)&{\cal B}_1(\lmd)\\
{\cal A}_{21}(\lmd)& {\cal A}_{22}(\lmd)&{\cal B}_2(\lmd)\\
{\cal C}_{1}(\lmd)& {\cal C}_{2}(\lmd)&{\cal D}(\lmd)
 \end{array}\right).
\end{eqnarray}
Around $\eta=0$:
\begin{eqnarray}
{\cal T}(\lmd)&=&1+\eta\hat{\cal T}(\lmd)+{\cal O}(\eta^2)
        \nonumber\\[3mm]
&=&1+\eta\left(\begin{array}{ccc}
\hat{\cal A}_{11}(\lmd)& \hat{\cal A}_{12}(\lmd)&\hat{\cal B}_1(\lmd)\\
\hat{\cal A}_{21}(\lmd)& \hat{\cal A}_{22}(\lmd)&\hat{\cal B}_2(\lmd)\\
\hat{\cal C}_{1}(\lmd)& \hat{\cal C}_{2}(\lmd)&\hat{\cal D}(\lmd)
 \end{array}\right)+{\cal O}(\eta^2). \label{limit-Topen}
\end{eqnarray}
Applying $\hat{\cal T}(\lmd=z_j),~~j=1,2,\cdots,N,$ to the vacuum state
(\ref{vacuum}), we have 
\begin{eqnarray}
&&\hat B_a(z_j )|0>=0,
~~~\hat C_a(z_j)|0>\not= 0
\nonumber \\
&&\hat D(z_j)|0>
   =\sum_{i=1}^N2(\coth(z_j-z_i)+\coth(z_j-z_i))|0>.\nonumber\\
&&\hat A_{ab}(\lambda )|0>=\left\{\begin{array}{l}
0\quad \mbox{for } \lmd=z_j,\mbox{ and } a\ne b \\ 
\sum_{i=1}^N(\coth(\lmd+z_i)+\coth(\lmd-z_i)) \quad
  \mbox{for }\lmd\ne z_j,\mbox{ and } a= b  
\end{array}\right.. \nonumber\\   \label{eigenv-o}
\end{eqnarray}
The Bethe state of the boundary system can still be taken as 
\begin{eqnarray}
\phi_b=\hat C_{d_1}(\mu _1)
\hat C_{d_2}(\mu _2)\cdots \hat C_{d_n}(\mu _n)|0>F^{d_1\cdots
d_n}. \end{eqnarray}
Write
\begin{eqnarray}
\hat{\cal A}(\lmd)_{ab}|_{\eta=0}
=\hat{\tilde{\cal A}}(\lmd)_{ab}|_{\eta=0}
 +\delta_{ab}{1\over\sinh(2\lmd)}{\cal D}(\lmd)_{\eta=0}, \label{tilde-A}
 \end{eqnarray} 
where
\begin{eqnarray}
\hat{\tilde{\cal A}}(\lmd)_{ab}|0>
      =\sum_{i=1}^N(\coth(\lmd+z_i)+\coth(\lmd-z_i))
       -{\delta_{ab}\over\sinh(2\lmd)}{\cal D}(\lmd)_{\eta=0}.
\end{eqnarray}
Then
\begin{eqnarray}
\hat t^b(\lmd=z_j)&\equiv& H_j^b
 ={d\over d\eta}\left(K^+{\cal T}(z_j)\right)_{\eta=0}\nonumber\\
&=&-\hat {\cal A}_{aa}(z_j)+\hat{\cal D}(z_j)
           \nonumber\\&&\mbox{}\quad
    -\left(k_a^+\right)'_{\eta=0}{\cal A}(z_j)_{\eta=0}
    -U{\cal D}(\lmd)_{\eta=0},   \label{t-b}
\end{eqnarray}
where $a=1,2$ and $U=2/\sinh(2z_j)$.
The last term in (\ref{t-b}) corresponds to the boundary condition.

We now find commutation relations between 
$\hat{\tilde{\cal A}}_{ab}(\lmd),\ \hat{\cal D}
(\lmd)$ and $\hat{\cal C}_{d_i}(\mu)$. After a tedious but direct
computation, we get
\begin{eqnarray}
\hat {\cal C}_{d_1}(\mu_1 )\hat {\cal C}_{d_2}(\mu_2 )
&=&-\hat {\cal {C}}_{c_2}(\mu_2 )\hat {\cal{C}}_{c_1}(\mu_1 ), 
\end{eqnarray}
\begin{eqnarray}
\hat {\cal{D}}(z_j )\hat {\cal{C}}_{d}(\mu )
&=&\hat{\cal C}_d(\mu)\hat{\cal D}(z_j)
-{\sinh(2z_j)\over\sinh(z_j-\mu)\sinh(z_j+\mu)}
 \hat{\cal {C}}_{d}(\mu )D(z_j)|_{\eta=0}
\nonumber \\
&&\mbox{}+
\frac {1}{\sinh(z_j -\mu)}
  \left( -E^-_j(d)\hat{\cal D}(\mu)
         +\hat{\cal {C}}_{d}(z_j){\cal D}(\mu)|_{\eta=0}
  \right)\nonumber \\
&&\mbox{}
-\frac{1}{\sinh(z_j+\mu)}
 \left(-E^-_j(d)\hat{\tilde{\cal A}}_{bd}(\mu )
 +\hat{\cal {C}}_b(z_j)\tilde{\cal A}_{bd}(\mu)|_{\eta=0}\right)
            \nonumber\\ 
&&\mbox{}+{2\coth(2\mu)\over\sinh(z_j-\mu)}
          E^-_j(d){\cal D}(\mu)|_{\eta=0}
 -{2\cosh(z_j+\mu)\over\sinh^2(z_j+\mu)}
     E^-_j(b)\tilde{\cal A}_{bd}(\mu)|_{\eta=0},\nonumber\\
\end{eqnarray}
\begin{eqnarray}
\hat{\tilde{\cal {A}}}_{a_1d_1}(z_j)\hat{\cal {C}}_{d_2}(\mu)
&=&
\hat{\cal {C}}_{d_2}(\mu )\hat{\tilde {\cal {A}}}_{a_1d_1}(z_j )
       \nonumber\\% &&\mbox{}
&&\mbox{}+
\left(r_{12}(z_j +\mu +\eta )_{a_1c_2}^{c_1b_2}
        r_{21}(z_j-\mu )_{b_1b_2}^{d_1d_2}\right)'_{\eta=0}
\hat{\cal {C}}_{c_2}(\mu )\tilde{\cal A}_{c_1b_1}(\lmd)|_{\eta=0}
\nonumber \\
&& \mbox{}+
{1\over\sinh(z_j-\mu )}\delta_{a_1b_2}\delta_{b_1d_1}\left(
    -E^-_j(b_1)\hat{\tilde{\cal{A}}}_{a_1d_2}(\mu)
    +\hat{\cal C}_{d_1}(z_j)\tilde{\cal{A}}_{a_1d_2}(\mu)|_{\eta=0}
 \right)
\nonumber \\
&&\mbox{}-
{1\over\sinh(z_j+\mu)}\delta_{a_1d_2}\delta_{b_2d_1}
   \left(-E^-_j(b_2)\hat{\cal D}(\mu)
         +\hat{\cal C}_{b_2}(z_j){\cal D}(\mu)|_{\eta=0}\right)
\nonumber \\
&&\mbox{}-
\left({\sinh(\eta)r_{12}(2z_j+\eta)^{b_2d_1}_{a_1b_1}\over
          \sinh(z_j-\mu)}\right)''_{\eta=0}
    E^-_j(b_1)\tilde{\cal A}_{b_2d_2}(\mu)|_{\eta=0}
\nonumber \\
&&\mbox{}+
\left({\sin(2\mu)sin(\eta)r_{12}(2z_j+\eta)_{a_1b_2}^{d_2d_1}\over
          \sinh(z_j+\mu+\eta)\sinh(2\mu+\eta)}
    \right)''_{\eta=0}
      E^-_j(b_2){\cal D}(\mu)|_{\eta=0},
\end{eqnarray}
\begin{eqnarray}
{\cal D}(z_j)_{\eta=0}\hat{\cal C}_d(\mu)
 &=&\hat{\cal C}_d(\mu){\cal D}(z_j)_{\eta=0}    
   +{1\over\sinh(z_j-\mu)}E^-_j(d){\cal D}(\mu)_{\eta=0}
             \nonumber\\&&\mbox{}
    -{1\over\sinh(z_j+\mu)}E^-_j(b)
      \tilde{\cal A}_{bd}(\mu)_{\eta=0},\\[3mm]
\tilde{\cal A}_{a_1d_1}(z_j)_{\eta=0}\hat{\cal C}_{d_2}(\mu)
 &=&\hat{\cal C}_{d_2}(\mu)\tilde{\cal A}_{a_1d_1}(z_j)_{\eta=0}    
   +{1\over\sinh(z_j-\mu)}E^-_j(d_1)\tilde{\cal A}_{a_1d_2}(\mu)_{\eta=0}
             \nonumber\\&&\mbox{}
    -{1\over\sinh(z_j+\mu)}\delta_{a_1d_2}E^-_j(d_1){\cal D}(\mu)_{\eta=0}
\end{eqnarray}

Then, applying (\ref{t-b}) to the Bethe
state and using the above commutation relations repeatedly, we 
obtain the off-shelled Bethe ansatz equations 
\begin{eqnarray}
H^b_j\phi^b
 &=&{dt(\lmd=z_j)\over d\eta}|_{\eta=0}\phi^b \nonumber\\
&=&E^b_j\phi^b
   -\sum_{\alpha=1}^nW^b(\mu_\alpha,z_j)f^b_\alpha E^-_j\phi^b_\alpha,
\end{eqnarray}
where 
\begin{eqnarray}
\phi^b&=&\prod_{\alpha=1}^n\left(\sum_{k=1}^N
   {2\sinh(\mu_\alpha)\cosh(z_k)\over
   \sinh(\mu_\alpha-z_k)\sinh(\mu_\alpha+z_k)}
           E^-_k(d_\alpha)\right)|0>F^{d_1\cdots d_n},\\
\phi^b_\alpha&=&
\prod_{\beta=1,\ne\alpha}^n\left(\sum_{k=1}^N
  {2\sinh(\mu_\beta)\cosh(z_k)\over
  \sinh(\mu_\beta-z_k)\sinh(\mu_\beta+z_k)}  
           E^-_k(d_\beta)\right)|0>F^{d_1\cdots d_n}.
\end{eqnarray}
\begin{eqnarray}
E^b_j&=&{2\over\sinh(z_j)}-2\sum_{k=1,\ne j}^N\left[
\coth(z_j+z_k)+\coth(z_j-z_k)\right]\nonumber\\
&&\mbox{}
+\sum_{\alpha=1}^n\left[
\coth(z_j+\mu_\alpha)+\coth(z_j-\mu_\alpha)\right]\\[3mm]
f^b_\alpha&=&
 -{2\over\sinh(2\mu_\alpha)}-\sum_{\beta=1,\ne\alpha}^n
 \left(\coth(\mu_\alpha-\mu_\beta)+\coth(\mu_\alpha+\mu_\beta)\right)
        \nonumber \\&&\mbox{}
+\sum_{k=1}^N
 \left(\coth(\mu_\alpha-z_k)+\coth(\mu_\alpha+z_k)\right)
+\Lambda^{(1)}_b(\mu_\alpha),\label{BA-b}\\[3mm]
W^b(\mu_\alpha,z_j)&=&{(-1)^{\alpha-1}2\sinh(z_j)\cosh(\mu_\alpha)
\over\sinh(z_j+\mu_\alpha)\sinh(z_j-\mu_\alpha)}.
\end{eqnarray}
Here $\Lambda^{(1)}_b$ is the eigenvalue of the nested transfer matrix
\begin{eqnarray}
&&t_b^{(1)}(\lmd)\nonumber\\
&=&{d\over d\eta}\left[
strK^{(1)+}r(\lmd+\mu_1+\eta)^{a_1e_1}_{a\ c_1}
           r(\lmd+\mu_2+\eta)^{a_2e_2}_{a_1c_2}\cdots
           r(\lmd+\mu_n+\eta)^{a_ne_n}_{a_{n-1} c_n}\right.
                  \nonumber\\&&\left.
K^{(1)}r_{21}(\lmd-\mu_n)^{b_{n-1}d_n}_{b_ne_n}\cdots
       r_{21}(\lmd-\mu_2)^{b_{1}d_2}_{b_2e_2}
       r_{21}(\lmd-\mu_n)^{b_{n}d_1}_{b_1e_1}\right],\label{t-bn}
\end{eqnarray}
where
$$K^{(1)+}=diag\left(e^{2\eta},1\right), \quad
K^{(1)}=(1-\sinh(\eta)/\sinh(2\lmd+\eta))\cdot id.$$
The detailed calculation of the nested eigenvalue is discussed 
in the appendix. Here we write the result
\begin{eqnarray}
\Lambda^{(1)}_b(\mu_\alpha)&=&{e^{-\mu_\alpha}\over\sinh(\mu_\alpha)}
 +\sum_{\beta=1,\ne\alpha}^n\left[
\coth(\mu_\alpha-\mu_\beta)+\coth(\mu_\alpha+\mu_\beta)\right)
           \nonumber\\&&\mbox{}
 -\sum_{\gamma=1}^m\left[
\coth(\mu_\alpha-\mu_\gamma^{(1)})
 +\coth(\mu_\alpha+\mu_\gamma^{(1)})\right),
\end{eqnarray}
where $\mu_\gamma^{(1)}$ satisfies the nested Bethe ansatz equation
\begin{eqnarray}
f^{b^{(1)}}_\gamma&=&\sum_{\beta=1}^n\left(\coth(\tilde\mu^{(1)}_\gamma
   +\tilde\mu_\beta)
   +\coth(\tilde\mu^{(1)}_\gamma-\tilde\mu_\beta)\right)\nonumber\\
&&\mbox{}-
2\sum_{\delta=1,\ne\gamma}^m
  \left(\coth(\tilde\mu^{(1)}_\gamma+\tilde\mu_\delta^{(1)})
        +\coth(\tilde\mu^{(1)}_\gamma-\tilde\mu_\delta^{(1)})\right)
\nonumber\\&=&0.
\end{eqnarray}
Substituting the nested eigenvalue into (\ref{BA-b}),
we obtain the Bethe ansatz equations for the boundary $t$-$J$
Gaudin model
\begin{eqnarray}
f^b_\alpha&=&-{e^{-\mu_\alpha}\over\cosh(\mu_\alpha)}
     +\sum_{k=1}^N
 \left(\coth(\mu_\alpha-z_k)+\coth(\mu_\alpha+z_k)\right)
           \nonumber\\&&\mbox{}
-\sum_{\gamma=1}^m
 \left(\coth(\mu_\alpha-\mu_\gamma^{(1)})
      +\coth(\mu_\alpha+\mu_\gamma^{(1)})\right)\nonumber\\
&=&0.
\end{eqnarray}

\sect{Super KZ equation in the boundary case}

As in the periodic case, the KZ equations are
\begin{equation}
\nabla_j\Psi = 0,~~~~\nabla_j=\kappa\frac{\partial}{\partial z_j}
  -H^b_j,~~~~j=1,2,\cdots,N,
\end{equation}
but now $H^b_j$ is the hamiltonian of the boundary $t$-$J$ Gaudin
model.  We make the transformation
\begin{eqnarray*}
H_j&\rightarrow &H_j-2/\sinh(2z_j)+2\sum_{k=1,\ne j}^N\left[
\coth(z_j+z_k)+\coth(z_j-z_k)\right],\\
E^b_j&\rightarrow& E^b_j
        -2/\sinh(2z_j)+2\sum_{k=1,\ne j}^N\left[
\coth(z_j+z_k)+\coth(z_j-z_k)\right]\\
&=&\sum_{\alpha=1}^n\left[
\coth(z_j+\mu_\alpha)+\coth(z_j-\mu_\alpha)\right]
\end{eqnarray*}
This transformation leaves invariant the form of  the
off-shell Bethe ansatz equations.

To construct $\Psi(z)$, we introduce a hypergeometric function
$\chi(z,\mu)$ which satisfies the following  equations
\begin{eqnarray}
&&\kappa\frac{\partial}{\partial z_j}\chi=E^b_j\chi, \nonumber\\
&&\kappa\frac{\partial}{\partial \mu_\alpha}\chi=f^b_\alpha\chi,
\end{eqnarray}
and the contraint $f^{b^{(1)}}_\gamma=0$.
Solving these two equations, one gets
\begin{eqnarray}
\chi(z,\mu)&=&
    \prod_{\alpha=1}^n
      \left(1+e^{-2\mu_\alpha}\right)^{1/\kappa}
    \prod_{\alpha=1}^n
    [\sinh(\mu_\alpha-\mu^{(1)}_\gamma)
     \sinh(\mu_\alpha-\mu^{(1)}_\gamma)]^{-1/\kappa}
\nonumber\\ &&\mbox{}\times
\prod_{\alpha=1}^n\prod_{j=1}^N
    [\sinh(z_j+\mu_\alpha)\sinh(z_j-\mu_\alpha)]^{1/\kappa},
\end{eqnarray}
where $\mu_\gamma^{(1)}$ satisfies the nested Bethe ansatz equation $
f^{b^{(1)}}_\gamma=0$.
With the help of $\chi(z,\mu)$, the function $\Psi(z)$ is given by
\begin{equation}
\Psi(z)=\oint_C\cdots\oint_C d\mu_1\cdots d\mu_M \chi(t,z)\phi(t,z),
\end{equation}  
where the integration path $C$ is a closed contour in the
Riemann surface such that the integrand resumes its initial value after
$t_\alpha$ has described it.
Substituting the expressions of $\nabla_j$ and $\Psi(z)$ into
({\ref{kz-elli}), we can show that the KZ equation is satisfied.
The proof is as follows
\begin{eqnarray}
\kappa\frac{\partial}{\partial z_j}\Psi(z)&=&
   \oint_C\cdots\oint_C d\mu_1\cdots d\mu_M \left(
       \kappa\frac{\partial \chi}{\partial z_j}\phi
       +\kappa\chi\frac{\partial \phi}{\partial z_j}\right)\nonumber\\ 
&=& \oint_C\cdots\oint_C dt_1\cdots d\mu_M \left(
       \chi E^b_j\phi
       +\kappa\chi\frac{\partial \phi}{\partial z_j}\right)\nonumber\\
&=&\oint_C\cdots\oint_C d\mu_1\cdots d\mu_M \left[
        \chi  H^b_j\phi+\chi\sum_{\alpha=1}^M
    W(\mu_\alpha,z_j)f_\alpha E^-_j(d_\alpha)\phi^b_\alpha\right.\nonumber\\
&&        +\left.\kappa\chi\sum_\alpha
\frac{\partial }{\partial \mu_\alpha}\left(
W(\mu_\alpha,z_j)\phi^b_\alpha E_j^-({d_\alpha})\right)\right]
                                       \nonumber\\ 
&=&\oint_C\cdots\oint_C d\mu_1\cdots d\mu_M \left[
        \chi H^b_j\phi\right.\nonumber\\
&&        +\left.\kappa\sum_\alpha
         \frac{\partial }{\partial \mu_\alpha}\left(\chi W(\mu_\alpha,z_j)
\phi^b_\alpha E_j^-({d_\alpha})\right)\right] \nonumber\\
&=&H^b_j\Psi,
\end{eqnarray}
where 
$$
W(\mu_\alpha,z)={(-1)^{\alpha-1}2\cosh(z_j)\sinh(\mu_\alpha)\over
       \sinh(\mu_\alpha+z_j)\sinh(\mu_\alpha-z_j)}.
$$

\vskip.3in
\noindent{\large\bf Acknowledgements}
\vskip.1in
This work has been financially supported by Australian Research
Council.

\vskip.3in
\appendix

\section{Details on the nested Bethe ansatz}

\subsection{Periodic case}

The nested Bethe ansatz was used to obtain eigenvalues of the
transfer matrix constructed from the following $r$-matrix:
\begin{eqnarray}
 r(\lmd)&=&{1\over \sinh(\lmd)}\left(\begin{array}{cccc}
\sinh(\lmd-\eta)&0&0&0 \\ 0&\sinh(\lmd)&-e^{-\lmd}\sinh(\eta)&0\\
0&-e^{\lmd}\sinh(\eta)&\sinh(\lmd)&0 \\ 0&0&0&\sinh(\lmd-\eta)
\end{array}\right).\nonumber\\
\end{eqnarray}

One can check that this $r$-matrix satisfies the unitarity and 
cross-unitarity relations
\begin{eqnarray}
&&r_{12}(\lmd)r_{21}(-\lmd)=-\sinh(\lmd+\eta)\sinh(\lmd+\eta)\cdot id.,
        \nonumber\\
&&r_{12}^{st_1}(2\eta-\lmd)r_{21}^{st_2}(\lmd)
       =\sinh(\lmd)\sinh(2\lmd-\eta)\cdot id.,
\end{eqnarray}
and the YBE
\begin{eqnarray}
r_{12}(\lmd-\mu)L^{(1)}_1(\lmd)L^{(1)}_2(\mu)
  =L^{(1)}_2(\mu)L^{(1)}_1(\lmd)r_{12}(\lmd-\mu),
\end{eqnarray}
where 
\begin{equation}
L^{(1)}_{j}(\lmd)={1\over\sinh(\lmd)}\left(\begin{array}{cc}
\sinh\left(\lmd-\eta{1+\sigma_j^z\over 2}\right)&-\sigma_j^-\sinh(\eta)\\
-\sigma_j^+\sinh(\eta)&\sinh\left(\lmd-\eta{1+\sigma_j^z\over 2}\right)
\end{array}\right)
\end{equation}
with $\sigma^\pm,\sigma^z$ being the usual Pauli matrices.
The monodromy matrix is 
\begin{eqnarray}
T^{(1)}(\lmd)&=&L^{(1)}_m(\lmd-\mu_m)\cdots L^{(1)}_2(\lmd-\mu_2)
               L^{(m)}_1(\lmd-\mu_1)\nonumber\\[3mm]
&\equiv&\left(\begin{array}{cc}
     A^{(1)}(\lmd)&B^{(1)}(\lmd)\\
     C^{(1)}(\lmd)&D^{(1)}(\lmd)\end{array}\right)\label{T-n},
\end{eqnarray}
which satisfies
\begin{eqnarray}
r_{12}(\lmd-\mu)T^{1)}_1(\lmd)T^{1)}_2(\mu)
 =T^{1)}_2(\mu)T^{1)}_1(\lmd)r_{12}(\lmd-\mu).\label{YBR-n}
\end{eqnarray}
The transfer matrix of the nested system is defined by
\begin{equation}
t^{(1)}(\lmd)=strT^{(1)}(\lmd).
\end{equation}
One can easily prove the commutativity of the transfer matrix: 
$[t^{(1)}(\lmd),t^{(1)}(\mu)]=0$. 

Expanding $r, L^{(1)},T^{(1)}$ and $T^{(1)}$ around $\eta=0$:
\begin{eqnarray}
r(\lmd)&=&1+\eta\hat r(\lmd)|_{\eta=0}+{\cal O}(\eta^2),
                                \nonumber\\
L^{(1)}(\lmd)&=&1+\eta\hat L^{(1)}(\lmd)|_{\eta=0}
                         +{\cal O}(\eta^2),\nonumber\\
T^{(1)}(\lmd)&=&1+\eta\hat T^{(1)}(\lmd)|_{\eta=0}
                         +{\cal O}(\eta^2),\nonumber\\
t^{(1)}(\lmd)&=&-2+\eta\hat t^{(1)}(\lmd)|_{\eta=0}
                        +{\cal O}(\eta^2),
\end{eqnarray}
where 
\begin{eqnarray}
\hat r(\lmd)&=&\left(\begin{array}{cccc}
-\coth(\lmd)&0&0&0 \\ 0&0&-e^{-\lmd}/\sinh(\lmd)&0\\
0&-e^{\lmd}/\sinh(\lmd)&0&0 \\ 0&0&0&-\coth(\lmd)
\end{array}\right).\nonumber\\[3mm]
\hat L^{(1)}_k(\lmd)&=&\left(\begin{array}{cc}
-{1+\sigma_k^z\over 2}\coth(\lmd) & -\sigma_k^-e^{-\lmd}/\sinh(\lmd)\\
-\sigma_k^+e^{\lmd}/\sinh(\lmd)&-{1-\sigma_k^z\over 2}\coth(\lmd)
\end{array}\right).\nonumber\\[3mm]
\hat T^{(1)}(\lmd)&=&\left(\begin{array}{cc}
     \hat A^{(1)}(\lmd)&\hat B^{(1)}(\lmd)\\
     \hat C^{(1)}(\lmd)&\hat D^{(1)}(\lmd)\end{array}\right),\nonumber\\[3mm]
t^{(1)}(\lmd)&=&str\left(\begin{array}{cc}
     \hat A^{(1)}(\lmd)&\hat B^{(1)}(\lmd)\\
     \hat C^{(1)}(\lmd)&\hat D^{(1)}(\lmd)\end{array}\right)
   =-\hat A^{(1)}(\lmd)-\hat D^{(1)}(\lmd).\label{trans-n}
\end{eqnarray}

By the nested YBE (\ref{YBR-n}), we find the following commutation
relations
\begin{eqnarray}
%\hat \sigma^-_k\hat C^{(1)}(\mu)
%    &=&C^{(1)}(\mu)\sigma^-_k,\\[3mm]
\hat C^{(1)}(\lmd)\hat C^{(1)}(\mu^{(1)})
    &=&\hat C^{(1)}(\mu^{(1)})\hat C^{(1)}(\lmd),
                \label{commu-n1}\\[3mm]
\hat A^{(1)}(\lmd)\hat C^{(1)}(\mu^{(1)})
  &=&-\coth(\lmd-\mu^{(1)})\hat C^{(1)}(\mu^{(1)})
   +\hat C^{(1)}(\mu^{(1)})\hat A^{(1)}(\lmd)\nonumber\\
  &&\mbox{}+{e^{-\lmd+\mu^{(1)}}\over\sinh(\lmd-\mu^{(1)})}
         \left( C^{(1)}(\lmd)\hat A^{(1)}(\mu^{(1)})
  +\hat C^{(1)}(\lmd)D^{(1)}(\mu^{(1)})\right)_{\eta=0},\nonumber\\
            \\[3mm]
\hat D^{(1)}(\lmd)\hat C^{(1)}(\mu^{(1)})
  &=&\coth(\lmd-\mu^{(1)})\hat C^{(1)}(\mu^{(1)})_{\eta=0}
   +\hat C^{(1)}(\mu^{(1)})\hat D^{(1)}(\lmd)\nonumber\\
  &&\mbox{}+{e^{-\lmd+\mu^{(1)}}\over\sinh(\lmd-\mu^{(1)})}
         \left(C^{(1)}(\lmd)\hat D^{(1)}(\mu^{(1)})
  +\hat C^{(1)}(\lmd)\tilde A^{(1)}(\mu^{(1)})\right)_{\eta=0}.
            \label{commu-n2}\nonumber\\
\end{eqnarray}

For the nest transfer matrix, we choose its vacuum state as
\begin{eqnarray}
|0>^{(1)}_k=\left(\begin{array}{c}0\\1\end{array}\right),\quad
|0>^{(1)}=\otimes^n_{k=1}|0>^{(1)}_k.
\end{eqnarray}
Applying the $T$-operator to the vacuum state, we have
\begin{eqnarray}
&&\hat A^{(1)}(\lmd)|0>^{(1)}=\hat B^{(1)}(\lmd)|0>^{(1)}=0,\quad
\hat C^{(1)}(\lmd)|0>^{(1)}\ne 0,\nonumber\\ &&
\hat D^{(1)}(\lmd)|0>^{(1)}=\sum_{i=1}^m\coth(\lmd-\mu_i)\quad 
     \lmd\ne \mu_k \ (k=1,2,\cdots,m).
\end{eqnarray}
The eigenvector of the nested transfer matrix (\ref{trans-n}) can be 
constructed from 
\begin{equation}
\phi^{(1)}=\hat C^{(1)}(\mu^{(1)}_1)\hat C^{(1)}(\mu^{(1)}_2)\cdots
           \hat C^{(1)}(\mu^{(1)}_m).
\end{equation}
Applying the nested transfer matrix to the above state and 
keeping in mind $\lmd=\mu_\alpha$,  
we obtain 
\begin{eqnarray}
\Lambda^{(1)}(\mu_\alpha)
 =\sum_{\beta=1,\ne\alpha}^n\coth(\mu_\alpha-\mu_\beta)
  -\sum_{\gamma=1}^m\coth(\mu_\alpha-\mu_\gamma^{(1)}),\quad 
\end{eqnarray}
where $\mu^{(1)}_1,\cdots, \mu^{(1)}_m$ satisfy the constraints
\begin{eqnarray}
f^{(1)}_\gamma\equiv 
   \sum_{\beta=1,\ne\alpha}^n\coth(\mu_\beta-\mu^{(1)}_\gamma)
   +2\sum_{\delta=1,\ne\gamma}^m
      \coth(\mu^{(1)}_\gamma-\mu^{(1)}_\delta)=0.
\end{eqnarray}

\subsection{Open boundary case}

In the open boundary case, we have boundary reflection and
dual relection equations
\begin{eqnarray}
&&r_{12}(\lmd-\mu)K^{(1)}_1(\lmd)r_{21}(\lmd+\mu)K^{(1)}_2(\mu)
=K^{(1)}_2(\mu)r_{12}(\lmd+\mu)K^{(1)}_1(\lmd)r_{21}(\lmd-\mu),
    \label{KE-n}     \nonumber\\  \\
&&r_{12}(\mu-\lmd)K^{(1)^+}_1(\lmd)M_1^{-1}
   r_{21}(2\eta-\lmd-\mu)K^{(1)^+}_2(\mu)M_2^{-1}\nonumber\\
&&\ \ =K^{(1)^+}_2(\mu)M_2^{-1}r_{12}(2\eta-\lmd-\mu)K^{(1)^+}_1(\lmd)
   M_1^{-1}r_{21}(\mu-\lmd).\label{DKE-n}
\end{eqnarray}
One can check $K^{(1)}=2\cosh(\eta+\lmd)\sinh(\lmd)/\sinh(2\lmd+\eta)$ and 
$K^{(1)^+}={\rm diag}(e^{2\eta},1)$ are solutions to the first and
second equations, respectively.

Using the nested monodromy matrix (\ref{T-n}), we define 
the double-row monodromy matrix for the open boundary system
\begin{eqnarray}
{\cal T}^{(1)}(\lmd)&\equiv& 
      T^{(1)}(\tilde\lmd)K^{(1)}(\lmd)T^{(1)^{-1}}(-\tilde\lmd)\nonumber\\
&=&\left(\begin{array}{cc}{\cal A}^{(1)}(\lmd)&{\cal B}^{(1)}(\lmd)\\
         {\cal C}^{(1)}(\lmd)&{\cal D}^{(1)}(\lmd)\end{array}\right),
    \label{T-nb}
\end{eqnarray}
where $T^{(1)}$ and $T^{(1)^{-1}}$ are defined by
\begin{eqnarray}
T^{(1)}_{aa_n}(\tilde\lmd)^{e_1\cdots e_n}_{c_1\cdots c_n}
&=&r(\tilde\lmd+\tilde\mu_1)^{a_1e_1}_{ac_1}
   r(\tilde\lmd+\tilde\mu_2)^{a_2e_2}_{a_1c_2}\cdots
   r(\tilde\lmd+\tilde\mu_n)^{a_ne_n}_{a_{n-1}c_n}\nonumber\\
&=&L_1^{(1)}(\tilde\lmd+\tilde\mu_1)L_2^{(1)}(\tilde\lmd+\tilde\mu_2)
   \cdots L_n^{(1)}(\tilde\lmd+\tilde\mu_n),\\[3mm]
T^{(1)^{-1}}(\tilde\lmd)
&=&r_{21}(\tilde\lmd-\tilde\mu_n)^{b_{n-1}d_n}_{b_ne_n}\cdots
   r(\tilde\lmd-\tilde\mu_2)^{b_1d_2}_{b_2e_2}
    r(\tilde\lmd-\tilde\mu_1)^{ad_1}_{b_{1}e_1}\nonumber\\
&=&L_n^{(1)^{-1}}(-\tilde\lmd+\tilde\mu_n)\cdots
   L_2^{(1)^{-1}}(-\tilde\lmd+\tilde\mu_2)
   L_1^{(1)^{-1}}(-\tilde\lmd+\tilde\mu_1),
\end{eqnarray}
respectively, and the $L$-operator takes the form
\begin{eqnarray}
L^{(1)}_k(\lmd)=\left(\begin{array}{cc}
b(\lmd)-(b(\lmd)-a(\lmd))e^{11}_k & -c_-(\lmd)e^{21}_k\\
-c_+(\lmd)e^{12}_k & b(\lmd)-(b(\lmd)-a(\lmd))e^{22}_k
\end{array}\right).
\end{eqnarray}
Let $\tilde\lmd=\lmd+\eta/2,\ \tilde\mu=\mu-\eta/2$, one sees
that the above definitions coincide with (\ref{t-bn}). 

The double-row monodromy matrix satisfies the reflection equation
\begin{equation}
r_{12}(\lmd-\mu){\cal T}^{(1)}_1(\lmd)
  r_{21}(\lmd+\mu){\cal T}^{(1)}_2(\mu)
 ={\cal T}^{(1)}_2(\mu)r_{12}(\lmd+\mu)
  {\cal T}^{(1)}_1(\lmd)r_{21}(\lmd-\mu),\label{RE-n}
\end{equation}
Thus, we can define the transfer matrix as 
\begin{equation}
t^{(1)}_b(\lmd)=strK^{(1)^+}{\cal T}^{(1)}(\lmd).
\end{equation}

Around $\eta=0$, we have the expansions
\begin{eqnarray}
{\cal T}^{(1)}(\lmd)&=&1+\eta\left(\begin{array}{cc}
\hat{\cal A}^{(1)}(\lmd)&\hat{\cal B}^{(1)}(\lmd)\\
\hat{\cal C}^{(1)}(\lmd)&\hat{\cal D}^{(1)}(\lmd)
        \end{array}\right)_{\eta=0}
+{\cal O}(\eta^2),\\[3mm]
t^{(1)}_b(\lmd)&=&-2+\eta\hat t^{(1)}_b(\lmd)_{\eta=0}
              +{\cal O}(\eta^2).
\end{eqnarray}
Write
\begin{eqnarray}
\hat{\cal A}^{(1)}(\lmd)|_{\eta=0}
  =\hat{\tilde{\cal A}}^{(1)}(\lmd)|_{\eta=0}
  -{e^{-2\lmd}\over\sinh(2\lmd)}{\cal D}^{(1)}(\lmd)|_{\eta=0}. \label{A-n}
\end{eqnarray}
The we find the commutation relations between $\hat t^{(1)}_b(\lmd)$ 
and $\hat{\cal C}^{(1)}(\mu^{(1)})$
 \begin{eqnarray}
\hat{\cal C}^{(1)}(\lmd)\hat{\cal C}^{(1)}(\mu^{(1)})
=\hat{\cal C}^{(1)}(\mu^{(1)})\hat{\cal C}^{(1)}(\lmd).          
\end{eqnarray}
\begin{eqnarray}
&&\hat{\cal D}^{(1)}(\lmd)\hat{\cal C}^{(1)}(\mu^{(1)})\nonumber\\
   &=&{\sinh(2\lmd)\over\sinh(\lmd+\mu^{(1)})\sinh(\lmd-\mu^{(1)})}
      \hat{\cal C}^{(1)}(\mu^{(1)}){\cal D}^{(1)}(\lmd)_{\eta=0}
     +\hat{\cal C}^{(1)}(\mu^{(1)})\hat{\cal D}^{(1)}(\lmd)\nonumber\\
    &&\mbox{}-{e^{(\lmd-\mu^{(1)})}\over\sinh(\lmd-\mu^{(1)})}\left(
      {\cal C}^{(1)}(\lmd)\hat{\cal D}^{(1)}(\mu^{(1)})
        +\hat{\cal C}^{(1)}(\lmd){\cal D}^{(1)}(\mu^{(1)})           
              \right)_{\eta=0}  \nonumber\\
    &&\mbox{}+{e^{(\lmd+\mu^{(1)})}\over\sinh(\lmd+\mu^{(1)})}\left(
      {\cal C}^{(1)}(\lmd)\hat{\tilde{\cal A}}^{(1)}(\mu^{(1)})
        +\hat{\cal C}^{(1)}(\lmd)
       \tilde{\cal A}^{(1)}(\mu^{(1)})\right)_{\eta=0},
\end{eqnarray}
\begin{eqnarray}
&&\hat{\tilde{\cal A}}^{(1)}(\lmd)\hat{\cal C}^{(1)}(\mu^{(1)})
           \nonumber\\
   &=&-{\sinh(2\lmd)\over\sinh(\lmd+\mu^{(1)})\sinh(\lmd-\mu^{(1)})}
           \hat{\cal C}^{(1)}(\mu^{(1)})
           {\tilde{\cal A}}^{(1)}(\lmd)_{\eta=0}
      +\hat{\cal C}^{(1)}(\mu^{(1)})\hat{\tilde{\cal A}}^{(1)}(\lmd)
                         \nonumber\\
    &&\mbox{}-{e^{-(\lmd+\mu^{(1)})}\over\sinh(\lmd+\mu^{(1)})}\left(
      {\cal C}^{(1)}(\lmd)\hat{\cal D}^{(1)}(\mu)+\hat{\cal C}^{(1)}(\lmd){\cal D}^{(1)}(\mu)\right)_{\eta=0}  \nonumber\\
    &&\mbox{}+{e^{-(\lmd-\mu^{(1)})}\over\sinh(\lmd-\mu^{(1)})}\left(
      {\cal C}^{(1)}(\lmd)\hat{\tilde{\cal A}}^{(1)}(\mu^{(1)})
      +\hat{\cal C}^{(1)}(\lmd)
       \tilde{\cal A}^{(1)}(\mu^{(1)})\right)_{\eta=0},
\end{eqnarray}
\begin{eqnarray}
{\cal D}^{(1)}(\lmd)\hat{\cal C}(\mu^{(1)})_{\eta=0}&=&
  \hat{\cal C}(\mu^{(1)}){\cal D}^{(1)}(\lmd)_{\eta=0}
 +{e^{\lmd-\mu^{(1)}}\over\sinh(\lmd-\mu^{(1)})}
     {\cal C}^{(1)}(\lmd){\cal D}^{(1)}(\mu^{(1)})_{\eta=0}
                                \nonumber\\&&\mbox{}
 -{e^{\lmd+\mu^{(1)}}\over\sinh(\lmd-\mu^{(1)})}
     {\cal C}^{(1)}(\lmd)\tilde{\cal A}^{(1)}(\mu^{(1)})_{\eta=0},
\end{eqnarray}
\begin{eqnarray}
\tilde{\cal A}^{(1)}(\lmd)\hat{\cal C}(\mu)_{\eta=0}&=&
  \hat{\cal C}(\mu)\tilde{\cal A}^{(1)}(\lmd)
 +{e^{-(\lmd-\mu)}\over\sinh(\lmd-\mu)}
   {\cal C}^{(1)}(\lmd)\tilde{\cal A}^{(1)}(\mu)_{\eta=0}
                      \nonumber\\&&\mbox{}
 -{e^{-(\lmd+\mu)}\over\sinh(\lmd-\mu)}
   {\cal C}^{(1)}(\lmd){\cal D}^{(1)}(\mu)_{\eta=0},
\end{eqnarray}
Substituting $K^{(1)}(\lmd)=2\cosh(\eta+\lmd)\sinh(\lmd)/\sinh(2\lmd+\eta)$ 
and $K^{(1)^+}(\lmd)={\rm diag}(e^{2\eta},1)$ into the nested transfer matrix,
one obtains the transfer matrix 
$\hat t^{(1)}_b(\tilde\lmd)$  of the $t$-$J$ Gaudin model
\begin{eqnarray}
\hat t^{(1)}_b(\tilde\lmd)&\equiv& {d\over d\eta}str
\left(\begin{array}{cc}e^{2\eta}&0\\0&1\end{array}\right)
\left(\begin{array}{cc} 
{\cal A}^{(1)}(\tilde\lmd)&{\cal B}^{(1)}(\tilde\lmd)\\
{\cal C}^{(1)}(\tilde\lmd)&{\cal D}^{(1)}(\tilde\lmd)
    \end{array}\right)_{\eta=0} \nonumber\\
&=&-\hat{\cal A}^{(1)}(\lmd)-\hat{\cal D}^{(1)}(\lmd)
   -2{\cal A}^{(1)}(\tilde\lmd)_{\eta=0}\nonumber\\
&=&-\hat{\tilde{\cal A}}^{(1)}(\lmd)
   -\hat{\cal D}^{(1)}(\lmd)
   -2{\tilde{\cal A}}^{(1)}(\tilde\lmd)_{\eta=0}
   -{e^{-2\tilde\lmd}\over\sinh(2\tilde\lmd)}
        {\cal D}^{(1)}(\tilde\lmd)_{\eta=0}.
\end{eqnarray}

As in the periodic case, define the vacuum state for the nested open 
boundary system
\begin{equation}
|0>^{(1)}=\otimes^n_{k=1}|0>_k^{(1)}
\end{equation}
with $|0>^{(1)}_k\equiv\left(\begin{array}{l}0\\1\end{array}\right).$
Then, applying the elements of $\hat{\cal T}(\lmd)$ to the vacuum, we obtain
\begin{eqnarray}
\hat{\cal B}^{(1)}(\lmd)|0>=0, \quad
\hat{\cal C}^{(1)}(\lmd)|0>\ne 0,
\end{eqnarray}
\begin{eqnarray}
\hat{\cal D}^{(1)}(\lmd)|0>
&=&-\left({1\over\sinh(2\lmd)}
+\sum_{i=1}^n(\coth(\lmd+\mu_i)
                  +\coth(\lmd-\mu_i)\right)|0>^{(1)}\nonumber\\
&&\quad\quad\quad\quad\quad       \lmd\ne \mu_k \ (k=1,2,\cdots,m).
\end{eqnarray}
Using the nested YBE (\ref{YBR-n}) and the transformation (\ref{A-n}), 
we obtain 
\begin{eqnarray}
\hat{\tilde{\cal A}}^{(1)}( \lmd)|0>^{(1)}
  =-{e^{-\lmd}\over\sinh(\lmd)}|0>^{(1)}.
\end{eqnarray}

The eigenvalues of the nested open boundary Gaudin system can be obtained
by  applying $\hat t^{(1)}_b(\lmd)$ to the eigenvector 
\begin{equation}
\phi^{(1)}_b
 =\hat {\cal C}^{(1)}(\mu^{(1)}_1)\hat {\cal C}^{(1)}(\mu^{(1)}_2)\cdots
           \hat {\cal C}^{(1)}(\mu^{(1)}_m)|0>.
\end{equation}
The result is 
\begin{eqnarray}
\Lambda^{(1)}(\tilde\lmd)
&=&{e^{(-\lmd})\over\sinh(\lmd)}
    +\sum_{i=1}^n\left(\coth(\lmd+\tilde\mu_i)
                     +\coth(\lmd-\mu_i)\right)\nonumber\\
   &&\mbox{}
   -\sum_{l=1}^m\left(\coth(\lmd+\mu_l^{(1)})
                     +\coth(\lmd-\mu_l^{(1)})\right),
\end{eqnarray}
where $\mu_l^{(1)}$ satisfy the constraints
\begin{eqnarray}
f^{(1)}_\gamma&=&
\sum_{i=1}^n\left(\coth(\mu^{(1)}_j+\mu_i)
                  +\coth(\mu^{(1)}_j-\mu_i)\right)\nonumber\\
&&\mbox{}-2\sum_{l=1,\ne j}^m\left(\coth(\mu^{(1)}_j+\mu_l^{(1)})
                  +\coth(\mu^{(1)}_j-\mu_l^{(1)})\right)\nonumber\\
 &=&0.
\end{eqnarray}

\end{document}